\documentclass[10pt, doublecolumn]{IEEEtran}
\usepackage{epsfig,latexsym}
\usepackage{float}
\usepackage{indentfirst}
\usepackage{amsmath}
\usepackage{bm}
\usepackage{amssymb}
\usepackage{times}
\usepackage{enumitem}

\usepackage{algorithm}
\usepackage[noend]{algpseudocode}
\usepackage{subfigure}
\usepackage{psfrag}
\usepackage{hyperref}
\usepackage{cite}
\usepackage{lastpage}
\usepackage{fancyhdr}
\usepackage{color} 
 \usepackage{amsthm}
\usepackage{bigints}
\usepackage{array}
\usepackage{booktabs}
\newcounter{problem}
\newcounter{save@equation}
\newcounter{save@problem}
\makeatletter

\begin{document}
\title{  \vspace{-1em} \huge{ Pinching-Antenna Assisted ISAC:  A CRLB Perspective }}

\author{ Zhiguo Ding, \IEEEmembership{Fellow, IEEE} \thanks{ 
  
\vspace{-2em}

Z. Ding is with the University
of Manchester, Manchester, M1 9BB, UK, and Khalifa University, Abu Dhabi, UAE.


  }\vspace{-2.6em}}
 \maketitle

\begin{abstract}
Recently, pinching antennas have attracted significant research interest due to their capability to reconfigure wireless channels as well as their array configuration flexibility. This letter focuses on how these features can be used to support integrated sensing and communications (ISAC) from the Cramér–Rao lower bound (CRLB) perspective. In particular, the CRLB achieved by pinching antennas is first derived and then compared to that of conventional antennas. The presented analytical and simulation results demonstrate that using pinching antennas can significantly reduce CRLB and, hence, enhance positioning accuracy.  In addition, this letter also reveals that the low-cost and reconfigurability features of pinching antennas can be utilized to realize flexible user-centric positioning.  
\end{abstract}\vspace{-0.2em}

 \vspace{-1.5em}
 \section{Introduction}
 Recently, pinching antennas have received significant attention from both academia and industry as a novel evolution of smart antennas, and offer three distinguished features \cite{pinching_antenna2,mypa}. One is their capability to create strong line-of-sight (LoS) links between the transceivers, which means that large-scale path losses and LoS blockage can be effectively mitigated by activating antennas close to users \cite{myblockage}.  The second feature is the reconfigurability of pinching-antenna systems, where the topology of a pinching-antenna array, e.g., the locations and the number of pinching antennas, can be flexibly adjusted. The third feature is their practicality, where DOCOMO's prototype shows that pinching antennas can be straightforwardly implemented in a low-cost manner\cite{pinching_antenna2}.  
 
In the literature, there already exists a large amount of work to demonstrate that the use of pinching antennas can significantly enhance the communication functionality of wireless networks. For example,  the fundamental issues of pinching antennas, such as antenna activation, the architecture of a pinching-antenna array, and the array gains, have been investigated in \cite{Kaidiactivation, Chongjunpa1,Chongjunpa1x}.  Antenna placement is key to realizing the full potential of pinching-antenna systems, where various designs and their impact on the system throughput have been investigated in \cite{yanqingpa,xidongpass1}. Channel estimation and beam training are crucial issues to pinching-antenna systems, and sophisticated designs using the flexibility features of pinching antennas have been developed in \cite{yuanwei1x,yuanwei2x}. For many resource allocation problems encountered in pinching-antenna systems, the use of conventional convex optimization leads to high computational complexity, which motivates the application of advanced learning methods \cite{luyangpass1,yuanwei3x}. The applications of pinching antennas to improve the uplink throughput and the security of communication networks have also been recently investigated in \cite{Tegosuppa,yuanwei5x}.

However, we note that the impact of pinching antennas on the sensing functionality of wireless networks has not yet been fully characterized in the literature, although the recent work in \cite{yarupass} demonstrated the importance of pinching antennas in integrated sensing and communication (ISAC) systems \cite{9737357}, which motivates this letter. In particular, in this letter, the Cramér–Rao lower bound (CRLB) is used as the performance metric to characterize the capability of pinching antennas for enhancing the positioning accuracy of ISAC networks. The CRLB achieved by pinching antennas is first derived in the letter, and then compared to conventional antennas. The presented analytical results reveal that the use of pinching antennas can ensure that users at different locations experience uniform positioning accuracy, whereas the use of conventional antennas can result in a significant disparity in accuracy among the users. In addition, the important properties of CRLB achieved by pinching antennas, such as the effects of antenna placement and the local maximums of CRLB, are also investigated.  Furthermore, this letter also reveals that the low-cost and reconfigurability features of pinching antennas can be utilized to realize flexible user-centric positioning.  

\vspace{-0.5em} 
\section{System Model}\label{section 2}
 Consider a pinching-antenna system that is deployed to provide ISAC services to $M$ single-antenna users, denoted by ${\rm U}_m$. Given the fact that there is already a rich literature on using pinching antennas to enhance communications, and also due to space limitations, the impact of pinching antennas on the sensing functionality is focused on in this letter. Without loss of generality, assume that $N$ pinching antennas are activated on $N_{\rm WG}$ waveguides. The location of the $n$-th pinching antenna is denoted by ${\boldsymbol \psi}_n^{\rm Pin}=(x_n^{\rm Pin},y_n^{\rm Pin},d_{\rm H})$, where $d_{\rm H}$ denotes the height of the waveguides.

 The service area is denoted by $\mathcal{A}$ and is assumed to be a rectangle with its two sides denoted by $D_{\rm W}$ and $D_{\rm L}$, respectively, and its center located at $(0,0,0)$. The users are assumed to be uniformly distributed in $\mathcal{A}$, and ${\rm U}_m$'s location is denoted by $\boldsymbol{\psi}_m= (x_m,y_m,0)$.     
 
 Denote the distance from the $n$-th pinching antenna to the $m$-th user by $d_{mn}$. Distance (range) estimates for the $m$-th user can be modeled as follows: \cite{4753258}
 \begin{align}
\hat{d}_{mn} = d_{mn} + w_{mn},
 \end{align}
 where $d_{mn}=\sqrt{\left(x_m-x_n^{\rm Pin}\right)^2+\left(y_m-y_n^{\rm Pin}\right)^2+d_{\rm H}^2}$, and $ w_{mn}$ is a zero-mean Gaussian distributed noise term whose variance is distance-dependent, i.e.,
 \begin{align}
 \sigma^2_{mn} =  K_E\left(\left(x_m-x_n^{\rm Pin}\right)^2+\left(y_m-y_n^{\rm Pin}\right)^2+d_{\rm H}^2\right),
 \end{align}
  $K_E$ denotes a system parameter decided by the range estimation environment. 
  \vspace{-0.3em}
 \section{Impact of Pinching Antennas on Positioning }

 \vspace{-0.3em}
 
 \subsection{CRLB Achieved by Pinching-Antenna Systems}
 
Without loss of generality, the impact of pinching antennas on ${\rm U}_m$'s localization is focused on. The joint probability density function (pdf) of $\hat{d}_{mn} $ conditioned on $ {d}_{mn} $,  $1\leq n \leq N$, is given by
 \begin{align}
 f (\hat{d}_{m1}, \cdots, \hat{d}_{mN} )
 =\prod^{N}_{n=1}\frac{1}{\sqrt{2\pi  \sigma^2_{mn} }}e^{-\frac{(\hat{d}_{mn} - d_{mn}  )^2}{2 \sigma^2_{mn} }},
 \end{align}
 whose log-likelihood function is given by
  \begin{align}
L\triangleq& \ln f (\hat{d}_{m1}, \cdots, \hat{d}_{mN} )
 =-\frac{N}{2}\ln(2\pi)   \\\nonumber &- \sum^{N}_{n=1}\ln \sigma_{mn}-\sum^{N}_{n=1} \frac{(\hat{d}_{mn} - d_{mn}  )^2}{2 \sigma^2_{mn} }. 
 \end{align}
Recall that the CRLB for $x_m$ and $y_m$ is given by
{\small \begin{align}
 \mathcal{E}\left\{\left(
\hat{x}_m- {x}_m
\right)^2+\left(
\hat{y}_m- {y}_m
\right)^2\right\} \geq \frac{1}{J^m_x} + \frac{1}{J^m_y}\triangleq{\rm CRB}_m,
\end{align}}
\hspace{-0.5em}where $\hat{x}_m$ and  $\hat{y}_m$ denote the estimates of $x_m$ and $y_m$, respectively, $  {J^m_x}=\mathcal{E}\left\{ -\frac{\partial^2 L}{\partial x_m^2}\right\}$ and $ {J^m_y}=\mathcal{E}\left\{ -\frac{\partial^2 L}{\partial y_m^2}\right\}$.

$\frac{\partial L}{\partial x_m}$ can be obtained as follows:
\begin{align}
\frac{\partial L}{\partial x_m}=&- \sum^{N}_{n=1}\frac{1}{\sigma_{mn}}\frac{\partial \sigma_{mn}}{\partial x_m} -\sum^{N}_{n=1} \frac{( {d}_{mn} - \hat{d}_{mn}  )}{ \sigma^2_{mn} }\frac{\partial d_{mn}}{\partial x_m} \\\nonumber &+\sum^{N}_{n=1} \frac{(\hat{d}_{mn} - d_{mn}  )^2}{\sigma^3_{mn} }\frac{\partial \sigma_{mn}}{\partial x_m} .
\end{align}
The expression of $\frac{\partial^2 L}{\partial x^2_m}$ is quite invovled; however, by using the fact that $\mathcal{E}\{\hat{d}_{mn} - d_{mn}  \}=0$ and following the steps similar to those in \cite{4753258}, the expectation of $\frac{\partial^2 L}{\partial x^2_m}$, i.e.,  $J^m_x$, can be obtained as follows:
\begin{align}
J^m_x=&   \sum^{N}_{n=1}\frac{\left(2K_E   +1 \right)}{ \sigma^2_{mn} }   \frac{\left(x_m-x_n^{\rm Pin}\right)^2}{ {\left(x_m-x_n^{\rm Pin}\right)^2+\left(y_m-y_n^{\rm Pin}\right)^2+d_{\rm H}^2}} . 
\end{align}     

$J^m_y$ can be obtained in a similar form, which means that the CRLB for estimating ${\rm U}_m$'s location can be expressed as follows: 
 \begin{align}\nonumber
{\rm CRB}_m =  &    \frac{ K_E }{\left(2K_E   +1 \right) }\left(\frac{1}{\sum^{N}_{n=1}  \frac{\left(x_m-x_n^{\rm Pin}\right)^2}{ \left(\left(x_m-x_n^{\rm Pin}\right)^2+\left(y_m-y_n^{\rm Pin}\right)^2+d_{\rm H}^2\right)^2} }
\right. \\ \label{crlb all}
&\left.+ \frac{1}{\sum^{N}_{n=1} \frac{\left(y_m-y_n^{\rm Pin}\right)^2}{\left( \left(x_m-x_n^{\rm Pin}\right)^2+\left(y_m-y_n^{\rm Pin}\right)^2+d_{\rm H}^2\right)^2} }\right). 
\end{align}
 \vspace{-1em}
\subsection{Performance Analysis Based on CRLB}
 \subsubsection{Performance Gain over Conventional Antennas}\label{subsectioncompare}
For the conventional-antenna benchmark, consider the use of a circular antenna array with its center located at $(0,0,0)$ and its radius being $\frac{\lambda}{4\sin\left(\frac{\pi}{N}\right)}$, which ensures that the minimal pairwise distance of the antennas is $\frac{\lambda}{2}$, where $\lambda$ denotes the wavelength.    By using the fact that the users are uniformly distributed within the service area, the performance gain of pinching antennas over conventional antennas can be evaluated as follows:
{\small  \begin{align}\label{gap}
 \Delta_{\rm CRB} = \int^{\frac{D_{\rm L}}{2}}_{-\frac{D_{\rm L}}{2}} \int^{\frac{D_{\rm W}}{2}}_{-\frac{D_{\rm W}}{2}} \left(
 {\rm CRB}_m-{\rm CRB}_m^{\rm Conv}
 \right) \frac{dy_m}{D_{\rm W}} \frac{dx_m}{D_{\rm L}},
 \end{align}}
\hspace{-1.2em} where ${\rm CRB}_m^{\rm Conv}$ can be obtained similarly to ${\rm CRB}_m$ by replacing the locations of the pinching antennas with those of the conventional antennas. The performance gain in \eqref{gap} can be straightforwardly evaluated via computer simulations, but a closed-form expression of $ \Delta_{\rm CRB} $ is difficult to obtain due to the factional expression of the CRLB. We note that the performance gain of pinching antennas over conventional antennas can also be illustrated by simply focusing on the user which is located at $\left(\frac{D_{\rm L}}{2}, 0,0\right)$. The use of conventional antennas can achieve the following CRLB:
  \begin{align}\nonumber
&{\rm CRB}_m^{\rm Conv} =    \left(\frac{1}{\sum^{N}_{n=1}  \frac{\left(\frac{D_{\rm L}}{2}-x_n^{\rm Conv}\right)^2}{ \left(\left(\frac{D_{\rm L}}{2}-x_n^{\rm Conv}\right)^2+\left( y_n^{\rm Conv}\right)^2+d_{\rm H}^2\right)^2} }
\right. \\ \nonumber
&\left.+ \frac{1}{\sum^{N}_{n=1} \frac{\left( y_n^{\rm Conv}\right)^2}{\left( \left(\frac{D_{\rm L}}{2}-x_n^{\rm Conv}\right)^2+\left( y_n^{\rm Conv}\right)^2+d_{\rm H}^2\right)^2} }\right)  \frac{ K_E }{\left(2K_E   +1 \right) }
\\&\overset{(a)}{\approx}\nonumber
     \frac{ K_E }{\left(2K_E   +1 \right) }\left(\frac{4\left(\frac{D_{\rm L}^2}{4} +d_{\rm H}^2\right)^2}{  N D_{\rm L}^2  }
 + \frac{\left( \frac{D_{\rm L}^2}{4} +d_{\rm H}^2\right)^2}{\sum^{N}_{n=1} \left( y_n^{\rm Conv}\right)^2 }\right)
 \\\label{similarity} & \overset{(b)}{\rightarrow} \infty,
\end{align}
where step (a) is due to the fact that the conventional antennas are clustered close to the center of the service area, and step (b) is due to the fact that $|y_n^{\rm Conv}|\rightarrow 0$ for conventional antennas, particularly for the case with high carrier frequencies (i.e., small wavelengths).  

On the other hand, pinching antennas do not suffer the singularity issue experienced by conventional antennas. For example, for the user located at $\left(\frac{D_{\rm L}}{2}, 0,0\right)$, the corresponding CRLB can be expressed as follows:
\begin{align}\nonumber
{\rm CRB}_m =  &    \frac{ K_E }{\left(2K_E   +1 \right) }\left(\frac{1}{\sum^{N}_{n=1}  \frac{\left(\frac{D_{\rm L}}{2}-x_n^{\rm Pin}\right)^2}{ \left(\left(\frac{D_{\rm L}}{2}-x_n^{\rm Pin}\right)^2+\left(y_n^{\rm Pin}\right)^2+d_{\rm H}^2\right)^2} }
\right. \\ 
&\left.+ \frac{1}{\sum^{N}_{n=1} \frac{\left(y_n^{\rm Pin}\right)^2}{\left( \left(\frac{D_{\rm L}}{2}-x_n^{\rm Pin}\right)^2+\left(y_n^{\rm Pin}\right)^2+d_{\rm H}^2\right)^2} }\right) .
\end{align}
For illustrative purposes, a simple upper bound on the CRLB achieved by pinching antennas can be obtained as follows:
\begin{align}\nonumber
{\rm CRB}_m  
\leq &    \frac{ K_E }{\left(2K_E   +1 \right) }\left( \frac{ \left(\left(\frac{D_{\rm L}}{2}-x_n^{\rm Pin}\right)^2+\left(y_n^{\rm Pin}\right)^2+d_{\rm H}^2\right)^2} {\left(\frac{D_{\rm L}}{2}-x_n^{\rm Pin}\right)^2}
\right. \\ \label{boundxx}
&\left.+ \frac{\left( \left(\frac{D_{\rm L}}{2}-x_n^{\rm Pin}\right)^2+\left(y_n^{\rm Pin}\right)^2+d_{\rm H}^2\right)^2}{\left(y_n^{\rm Pin}\right)^2}\right),
\end{align}
where $n$ is an arbitrary integer between $1$ and $N$. Because of the diverse locations of the $N$ pinching antennas, it is always possible to find  $n\in\{1, \cdots, N\}$ which yields a finite value for the upper bound shown in \eqref{boundxx}, i.e., the CRLB achieved by pinching antennas is always bounded.
 
 {\it Remark 1:} Unlike conventional antennas which can cause noticeable accuracy variations between users, the carried-out case study shows that pinching antennas have the ability to offer uniform positioning accuracy between the users.  
  \vspace{-0.2em}
 \subsubsection{Flexible User-Centric Positioning} 
Due to their low-cost and reconfigurability features, the locations of pinching antennas can be tailored to a serving user for realizing flexible user-centric positioning. To facilitate the performance analysis, the association between the pinching antennas and the waveguides is required. Without loss of generality, assume that there are $\tilde{N}=\frac{N}{N_{\rm WG}}$ pinching antennas on each waveguide. Denote the location of the $n$-th antenna on the $i$-th waveguide by ${\boldsymbol \psi}_{in}^{\rm Pin}=(x_{in}^{\rm Pin},y_{in}^{\rm Pin},d_{\rm H})$. Furthermore, assume that the antennas are equally spaced, and define  $\Delta_x=\left|
x_{in}^{\rm Pin}-x_{im}^{\rm Pin}
\right|$ and $\Delta_y=\left|
x_{in}^{\rm Pin}-x_{jn}^{\rm Pin}
\right|$, $m\neq n$ and $i\neq j$. 

For illustrative purposes, assume that all $N$ pinching antennas are activated in a square area with ${\rm U}_m$ at its center, where $\tilde{N}=N_{\rm WG}$ and $\Delta_x=\Delta_y=\Delta$. This assumption is made to facilitate the performance analysis, and more practical setups will be considered in the simulation section. Define $\bar{N}=\frac{\tilde{N}}{2}$, and without loss of generality, assume that $\bar{N}$ is an even number. 

With these assumptions, the CRLB in \eqref{crlb all} can be simplified as follows:
\begin{align}\nonumber
{\rm CRB}_m =  &    \frac{ \frac{ K_E \Delta^2 }{4\left(2K_E   +1 \right) }}{\sum^{\bar{N}}_{i=1}\sum^{\bar{N}}_{n=1}  \frac{\left(n-\frac{1}{2} \right)^2}{\beta_{ni}^2} } + \frac{ \frac{ K_E \Delta^2 }{4\left(2K_E   +1 \right) }}{\sum^{\bar{N}}_{i=1}\sum^{\bar{N}}_{n=1}\frac{\left(i-\frac{1}{2} \right)^2}{\beta_{ni}^2} } ,
\end{align}
 where $=\left(n-\frac{1}{2} \right)^2+\left(i-\frac{1}{2}  \right)^2+\frac{d_{\rm H}^2}{\Delta^2}$.
The above CRLB can be used to design the antenna placement, i.e., the optimal choice of $\Delta$ for minimizing the CRLB.  Computer simulations can be used to verify that $\frac{\partial^2 {\rm CRB}_m}{\partial \Delta^2}>0$, i.e., $ {\rm CRB}_m$ is a convex function of $\Delta$, and hence convex optimization solvers can be used to find the optimal solution of $\Delta$ efficiently. To obtain an insightful understanding of the optimal choice of $\Delta$, a special case with $N=4$ is focused on in the following. We note that this special case is important in practice, given the fact that using a small number of antennas is helpful in reducing system overhead. For the case with $N=4$, the CRLB  can be simplified as follows:
  \begin{align}\label{simplified crlb} 
& {\rm CRB}_m = \frac{ 2K_E \Delta^2 }{\left(2K_E   +1 \right) }\left(\frac{1}{2}+\frac{d_{\rm H}^2}{\Delta^2}\right)^2  ,
\end{align}
whose first-order derivative is given by
  \begin{align}\label{first orderxx} 
&\frac{\partial {\rm CRB}_m}{\partial \Delta} = \frac{ 4K_E }{\left(2K_E   +1 \right) }\left(\frac{1}{2}\Delta+\frac{d_{\rm H}^2}{\Delta}\right)\left(\frac{1}{2} -\frac{d_{\rm H}^2}{\Delta^2}\right).
\end{align}
The second-order derivative of ${\rm CRB}_m$ is given by
  \begin{align}\label{second orderxx} 
\frac{\partial^2 {\rm CRB}_m}{\partial \Delta^2}  
=& \frac{ 4K_E }{\left(2K_E   +1 \right) }\left(\frac{1}{4}   + 3\frac{d_{\rm H}^4}{\Delta^4}   
\right)>0,
\end{align}
which means that $ {\rm CRB}_m$ is a convex function of $\Delta$. Therefore, the optimal solution of $\Delta$ for minimizing the CRLB for the special case with $N=4$ is given by 
\begin{align}\label{optimal}
\Delta^*=\sqrt{2}d_{H}. 
\end{align}
{\it Remark 2:} An intuition is that the CRLB is minimized if all the antennas are placed as close to the user as possible, i.e., $\Delta^*\rightarrow 0$ (or $\frac{\lambda}{2}$ to avoid antenna coupling). \eqref{optimal} shows that this intuition is wrong, where the optimal antenna spacing is a function of the height of the waveguides.  

 \vspace{-0.2em}
\subsubsection{Local-Maximum Property of CRLB}\label{subsection local}
 In the proximity of each pinching antenna, ${\boldsymbol \psi}_n^{\rm Pin}$, there exists a local maximum of ${\rm CRB}_m$ shown in \eqref{crlb all}.  
This local-maximum property can be revealed by studying $\frac{\partial {\rm CRB}_m}{\partial x_m}$ and $\frac{\partial {\rm CRB}_m}{\partial y_m}$. Without loss of generality, $\frac{\partial {\rm CRB}_m}{\partial x_m}$ is focused, and can be expressed as follows: 
 \begin{align}\label{gammaall}
\frac{\partial {\rm CRB}_m}{\partial x_m}= &   \frac{ K_E }{\left(2K_E   +1 \right) }\left(- \frac{1}{\gamma_1^2} \left[\gamma_2  - \gamma_3  \right] + \frac{1}{\gamma_4^2} \gamma_5\right), 
\end{align}
where $d_{mn}^2=\left(x_m-x_n^{\rm Pin}\right)^2+\left(y_m-y_n^{\rm Pin}\right)^2+d_{\rm H}^2$, $\gamma_1= \sum^{N}_{n=1}  \frac{\left(x_m-x_n^{\rm Pin}\right)^2}{ d_{mn}^4}  $, $\gamma_2= \sum^{N}_{n=1}  \frac{2\left(x_m-x_n^{\rm Pin}\right)}{ d_{mn}^4} $, $\gamma_3=\sum^{N}_{n=1}  \frac{4\left(x_m-x_n^{\rm Pin}\right)^3}{ d_{mn}^6}$, $\gamma_4=\sum^{N}_{n=1} \frac{\left(y_m-y_n^{\rm Pin}\right)^2}{d_{mn}^2}$, and $\gamma_5=\sum^{N}_{n=1} \frac{4 \left(x_m-x_n^{\rm Pin}\right)\left(y_m-y_n^{\rm Pin}\right)^2}{d_{mn}^6} $.

Without loss of generality, assume that ${\rm U}_{ m}$ is in the proximity of the first pinching antenna on the first waveguide, i.e.,  $x_m= x_{11}^{\rm Pin}+\delta_x$ and $y_m= y_{11}^{\rm Pin}+\delta_y$, where $\delta_x\rightarrow 0$ and $\delta_y\rightarrow 0$.  In this case, $\gamma_1$ in \eqref{gammaall} can be approximated as follows:
\begin{align}
\gamma_1\approx &    \sum^{\frac{N}{\tilde{N}}}_{i=1} \sum^{\tilde{N}-1}_{n=1}  \frac{n^2\Delta_x^2 }{ \left( n^2\Delta_x^2+(i-1)^2\Delta_y^2 +d_{\rm H}^2\right)^2},
\end{align}
where the terms at the order of $\delta_x^2$ are omitted. Similarly, by omitting the terms 
of $\delta_x^2$, $\gamma_2$ can be approximated as follows:
\begin{align}
\gamma_2 \approx &\sum^{\frac{N}{\tilde{N}}}_{i=1}    \frac{2\delta_x}{ \left(\delta^2+(i-1)^2\Delta_y^2+d_{\rm H}^2\right)^2}  \\\nonumber & - \sum^{\frac{N}{\tilde{N}}}_{i=1} \sum^{\tilde{N}-1}_{n=1}  \frac{2n \Delta_x  }{ \left( n^2\Delta_x^2+(i-1)^2\Delta_y^2 +d_{\rm H}^2\right)^2}.
\end{align}
Similarly, $\gamma_3$, $\gamma_4$ and $\gamma_5$ can be approximated as follows:
\begin{align}
\gamma_3  \approx&-  \sum^{\frac{N}{\tilde{N}}}_{i=1} \sum^{\tilde{N}-1}_{n=1}  \frac{4n^3 \Delta_x^3  }{ \left( n^2\Delta_x^2+(i-1)^2\Delta_y^2 +d_{\rm H}^2\right)^3},
\\\gamma_4 
 \approx &  \sum^{\frac{N}{\tilde{N}}-1}_{i=1} \sum^{\tilde{N}}_{n=1}  \frac{i^2\Delta^2_y}{\left(  (n-1)^2\Delta_x ^2+ i^2\Delta^2_y +d_{\rm H}^2\right)^2},\\
 \gamma_5 
\approx &  -\sum^{\frac{N}{\tilde{N}}-1}_{i=1} \sum^{\tilde{N}-1}_{n=1}  \frac{n\Delta_xi^2\Delta_y^2}{\left( n^2\Delta^2_x+ i\Delta^2_y +d_{\rm H}^2\right)^3} .
\end{align}

To facilitate the analysis of this local-maximum property of CRLB, assume that $\Delta_x=\Delta_y=  \Delta\gg d_{\rm H}$ and $\tilde{N}=\frac{N}{\tilde{N}}$, which means that $\gamma_1=\gamma_3$, and hence the CRLB can be simplified as follows:
  \begin{align}\nonumber
&\frac{\partial {\rm CRB}_m}{\partial x_m}\approx      \frac{\gamma_1 K_E }{\left(2K_E   +1 \right) } \left[-\sum^{\frac{N}{\tilde{N}}}_{i=1}    \frac{2\delta_x}{ \left(\delta^2+(i-1)^2\Delta^2+d_{\rm H}^2\right)^2} \right. \\  & \left. +\sum^{\frac{N}{\tilde{N}}}_{i=1} \sum^{\tilde{N}-1}_{n=1}  \frac{2n \Delta  }{ \bar{\beta}_{ni}^2} - \sum^{\frac{N}{\tilde{N}}}_{i=1} \sum^{\tilde{N}-1}_{n=1}  \frac{4n^3 \Delta^3  }{ \bar{\beta}_{ni}^3}  - \sum^{\frac{N}{\tilde{N}}-1}_{i=1} \sum^{\tilde{N}-1}_{n=1}  \frac{4n^3\Delta^3}{\bar{\beta}_{ni}^3} \right],\nonumber
\end{align}
where $\bar{\beta}_{ni}=(n^2+(i-1)^2)\Delta^2 +d_{\rm H}^2$.

Note that if $i=\frac{N}{\tilde{N}}$, $  \sum^{\tilde{N}-1}_{n=1}  \frac{4n^3\Delta^3}{\left( (n^2+ (i-1)^2)\Delta^2 +d_{\rm H}^2\right)^3} $ is an insignificant term, which means that the CRLB can be further simplified as follows: 
 \begin{align}\nonumber
&\frac{\partial {\rm CRB}_m}{\partial x_m}\approx      \frac{\gamma_1 K_E }{\left(2K_E   +1 \right) } \left[-\sum^{\frac{N}{\tilde{N}}}_{i=1}    \frac{2\delta_x}{ \left(\delta^2+(i-1)^2\Delta^2+d_{\rm H}^2\right)^2} \right. \\  & \left. +2 \Delta \sum^{\frac{N}{\tilde{N}}}_{i=1} \sum^{\tilde{N}-1}_{n=1}  \frac{n   \left( ( (i-1)^2-3n^2)\Delta^2 +d_{\rm H}^2\right)   }{ \left( (n^2+(i-1)^2)\Delta^2 +d_{\rm H}^2\right)^3}   \right]. 
\end{align}
 
For the case with $\delta_x=0$, i.e., the user is located right underneath of the pinching antenna at ${\boldsymbol \psi}_{11}^{\rm Pin}$, by using the assumption that  $\Delta\gg d$, the CRLB can be expressed as follows:  
 \begin{align}
\frac{\partial {\rm CRB}_m}{\partial x_m}\approx &     \frac{\gamma_1 K_E }{\left(2K_E   +1 \right) } 
\frac{2}{\Delta^3}  \gamma_6,
\end{align}
where $\gamma_6= \sum^{\frac{N}{\tilde{N}}}_{i=1} \sum^{\tilde{N}-1}_{n=1}   \frac{(i-1)^2-3n^2     }{  (n^2+(i-1)^2)^3 }  $.  We note that the terms of $\gamma_6$ decay rapidly by increasing $n$ and $i$, i.e., $\gamma_6$ can be approximated by keeping the dominant negative term ($n=1$ and $i=1$) and the dominant positive term ($n=1$ and $i=3$), i.e., $
\gamma_6\approx -3+\frac{1}{125}$,
which means $\frac{\partial {\rm CRB}_m}{\partial x_m}\leq 0$ for the case with $\delta_x=0$. For the case of $\delta_x\neq 0$, the CRLB can be approximated as follows:
 \begin{align}\nonumber
\frac{\partial {\rm CRB}_m}{\partial x_m}\approx  &    \frac{\gamma_1 K_E }{\left(2K_E   +1 \right) } \left[-    \frac{2\delta_x}{  d_{\rm H}^4}+\frac{2}{\Delta^3}  \gamma_6  \right].
\end{align}
 Due to the assumption of  $\Delta\gg d_{\rm H}$,  the term $    \frac{2\delta_x}{  d_{\rm H}^4}$ is dominant, and hence $\frac{\partial {\rm CRB}_m}{\partial x_m}>0$ if $\delta_x<0$.  In summary, $\frac{\partial {\rm CRB}_m}{\partial x_m}<0$ if the user's location is $( x_{11}^{\rm Pin}, y_{11}^{\rm Pin},0)$, and $\frac{\partial {\rm CRB}_m}{\partial x_m}>0$ if the user's location is $( x_{11}^{\rm Pin}+\delta_x, y_{11}^{\rm Pin}+\delta_y,d_{\rm H})$. A similar conclusion can be established to $\frac{\partial {\rm CRB}_m}{\partial y_m}$, which means that there exists a local maximum for the CRLB around ${\boldsymbol \psi}_n^{\rm Pin}$. 
 
 {\it Remark 3:} The local maximum property of the CRLB indicates an interesting conflict between the communication and sensing functionalities of pinching antennas. In particular, placing a pinching antenna directly above the user might increase the user's data rate but also degrade positioning accuracy. In other words, this local maximum property reveals the importance of antenna placement in pinching-antenna assisted ISAC networks.
 
\section{Numerical Studies} 
In this section, computer simulation results are presented to demonstrate the impact of pinching antennas on the positioning accuracy, where $K_E=0.01$, $D_{\rm W}=10$ m and $D_{\rm L}=40$ m, unless stated otherwise. 
     \begin{figure}[t]\centering \vspace{-0.2em}
    \epsfig{file=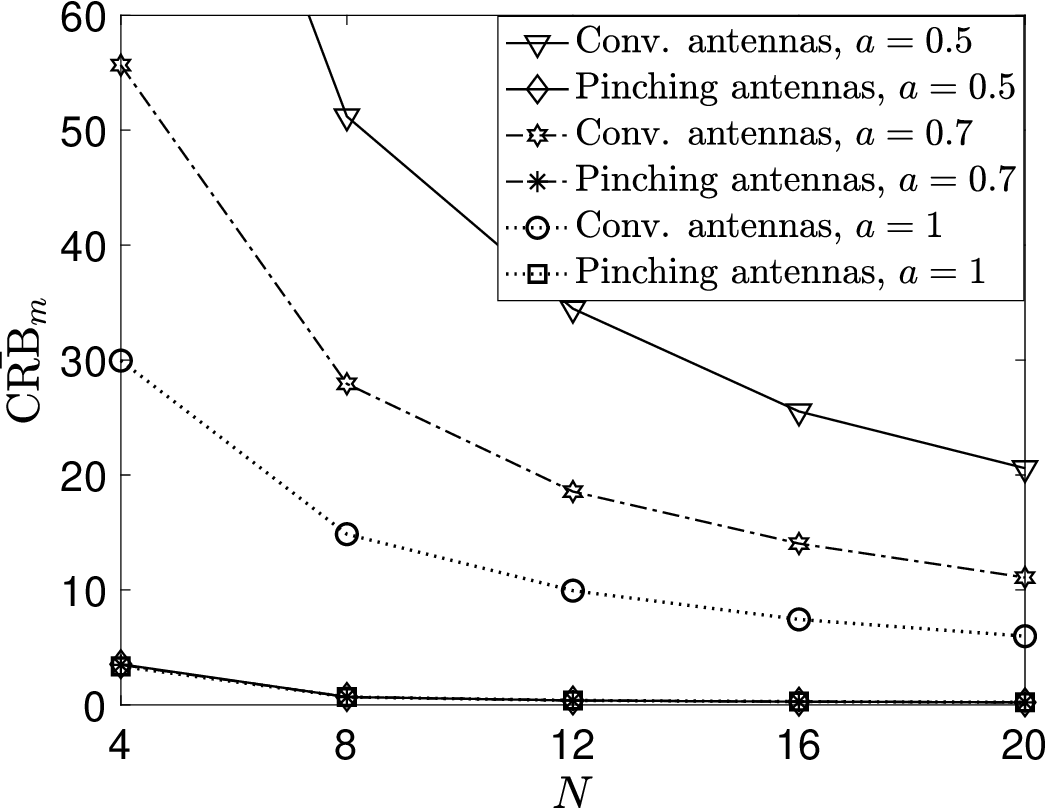, width=0.33\textwidth, clip=}\vspace{-0.5em}
\caption{Averaged CRLBs, $\bar{ {\rm CRB}_m}$, achieved by the considered antenna systems, where $N_{\rm WG}=2$ and  $d=3$ m. For the pinching-antenna system, on each waveguide, there are $\frac{N}{N_{\rm WG}}$ antennas, which are equally spaced. Due to the singularity issue experienced by conventional antennas discussed in Section \ref{subsectioncompare}, users are assumed to be excluded from a square area with its side being $a$ and its center at the origin.  
  \vspace{-1em}    }\label{fig0}   \vspace{-0.5em} 
\end{figure}

In Fig. \ref{fig0}, the averaged CRLBs achieved by the conventional and pinching-antenna systems are shown as functions of the number of antennas, where ${\rm U}_m$ is assumed to be uniformly deployed in the service area. Because the conventional-antenna system suffers the singularity issue discussed in Section \ref{subsectioncompare}, it is assumed that ${\rm U}_m$ cannot be located in a square area with its side being $a$ and its center at the origin. As can be seen from Fig. \ref{fig0}, the use of pinching antennas yields a significant performance gain over conventional antennas, regardless of the choices of $N$ and $a$.

   \begin{figure}[t] \vspace{-0.2em}
\begin{center}
\subfigure[Conventional Antennas]{\label{fig1a}\includegraphics[width=0.33\textwidth]{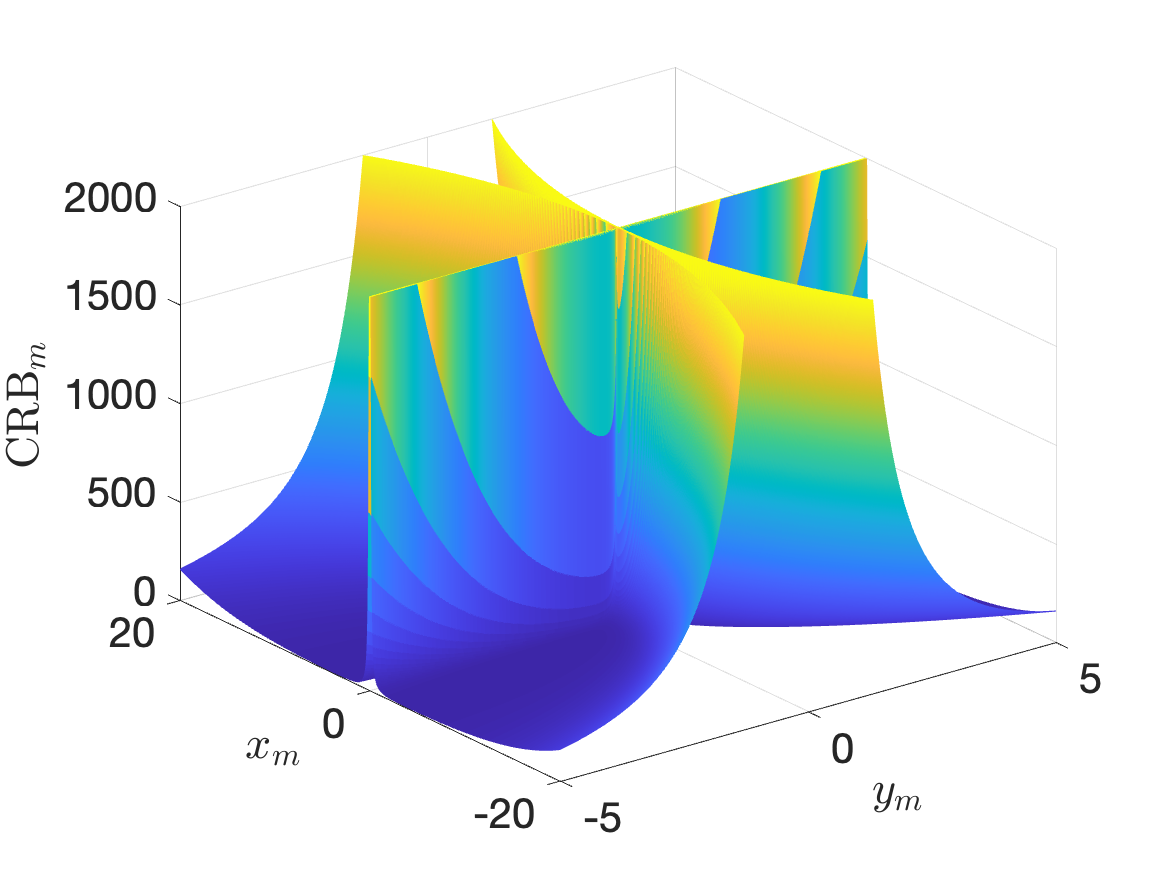}} 
\subfigure[Pinching Antennas]{\label{fig1b}\includegraphics[width=0.33\textwidth]{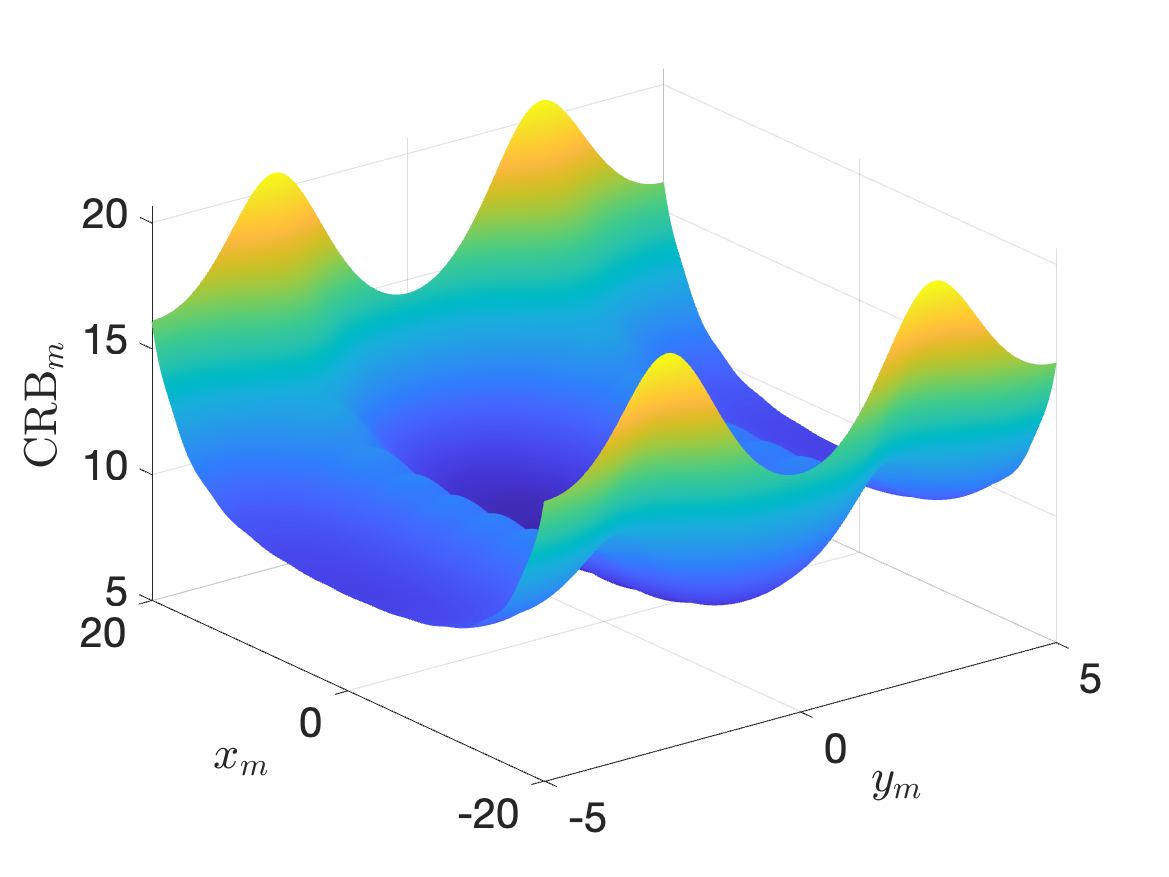}} \vspace{-1.5em}
\end{center}
\caption{CRLBs achieved by the considered antenna systems. $N=20$, $N_{\rm WG}=2$ and $d=3$ m. On each waveguide, there are $\frac{N}{N_{\rm WG}}$ antennas, which are equally spaced.  \vspace{-1em}.   }\label{fig1}\vspace{-1.2em}
\end{figure}

Fig. \ref{fig1} is provided to highlight the fact that a user's positioning accuracy depends on its location.  On the one hand, Fig. \ref{fig1a} shows that for conventional antennas, a user can experience extremely poor positioning accuracy if it is located far away from the center of the service area, which confirms the analytical results shown in \eqref{similarity}.  On the other hand, Fig. \ref{fig1b} shows that the use of pinching antennas ensures reasonably accurate positioning, regardless of whether the user is at the center or the edge of the service area. This also means that for the multi-user scenario, using pinching antennas can ensure fairness for the users' positioning accuracy.  We note that in Fig. \ref{fig1b}, local maximums are clearly visible in the proximity of the pinching antennas, which confirms the analysis shown in Section \ref{subsection local}. 
   \begin{figure}[t] \vspace{-2.2em}
\begin{center}
\subfigure[Positioning with a focal point at $\left(-\frac{D_{\rm L}}{4},0,0\right)$]{\label{fig2a}\includegraphics[width=0.33\textwidth]{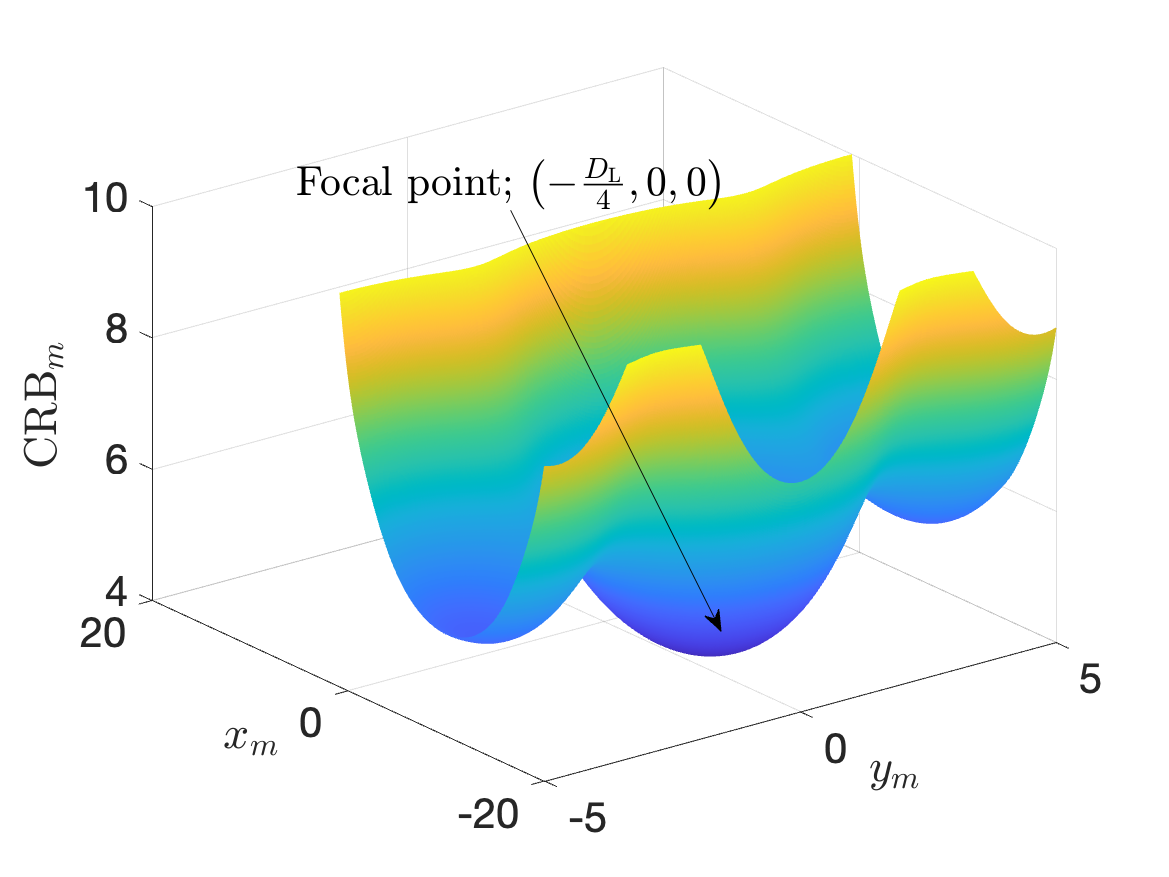}} 
\subfigure[Positioning with a focal point at $\left(\frac{D_{\rm L}}{4},0,0\right)$]{\label{fig2b}\includegraphics[width=0.33\textwidth]{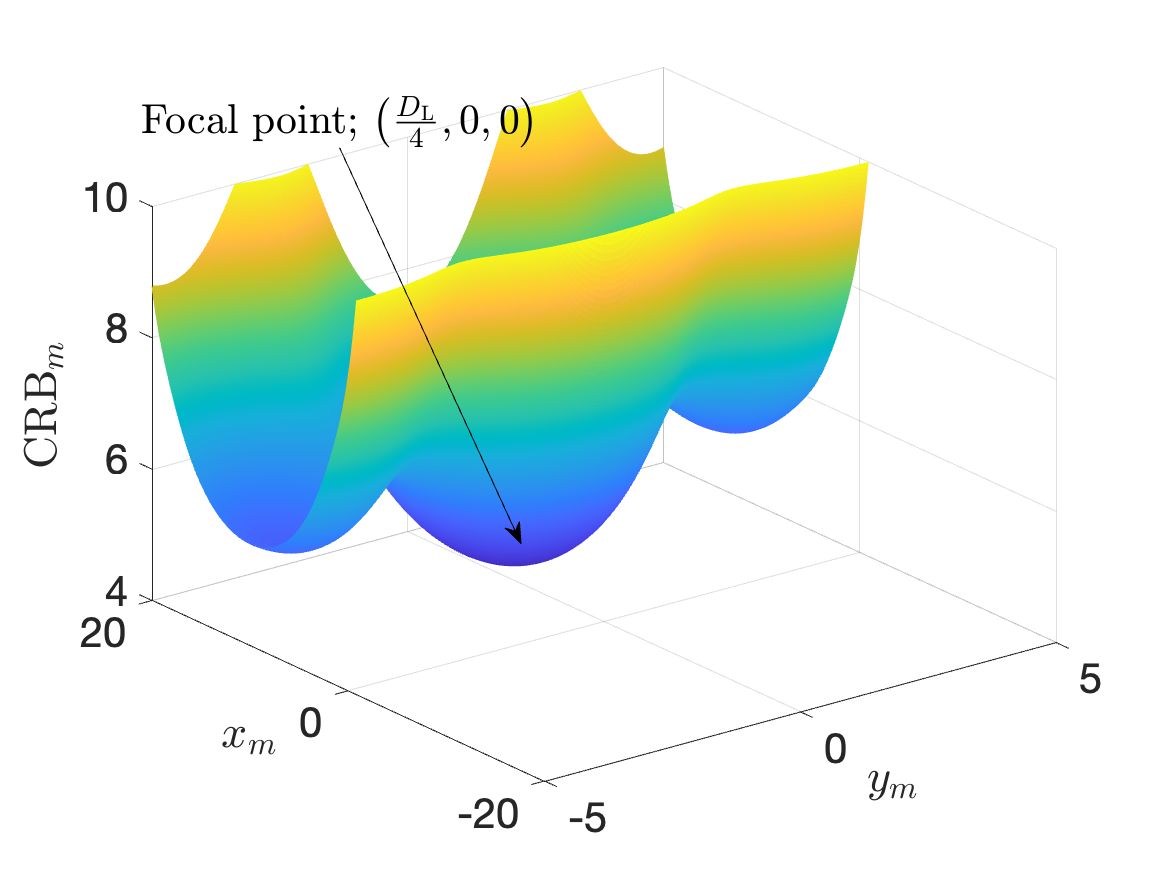}} \vspace{-1.5em}
\end{center}
\caption{Using pinching antennas to achieve flexible user-centric positioning.     $N=20$,  $N_{\rm WG}=2$ and $d=3$ m. On each waveguide, there are $\frac{N}{N_{\rm WG}}$ antennas, which are equally spaced in a segment with its length being $\frac{D_{\rm L}}{2}$ and its center at the focal points shown in the figures.    \vspace{-1em} }\label{fig2}\vspace{-1.2em}
\end{figure}

\begin{figure}[t]\centering \vspace{-2.2em}
    \epsfig{file=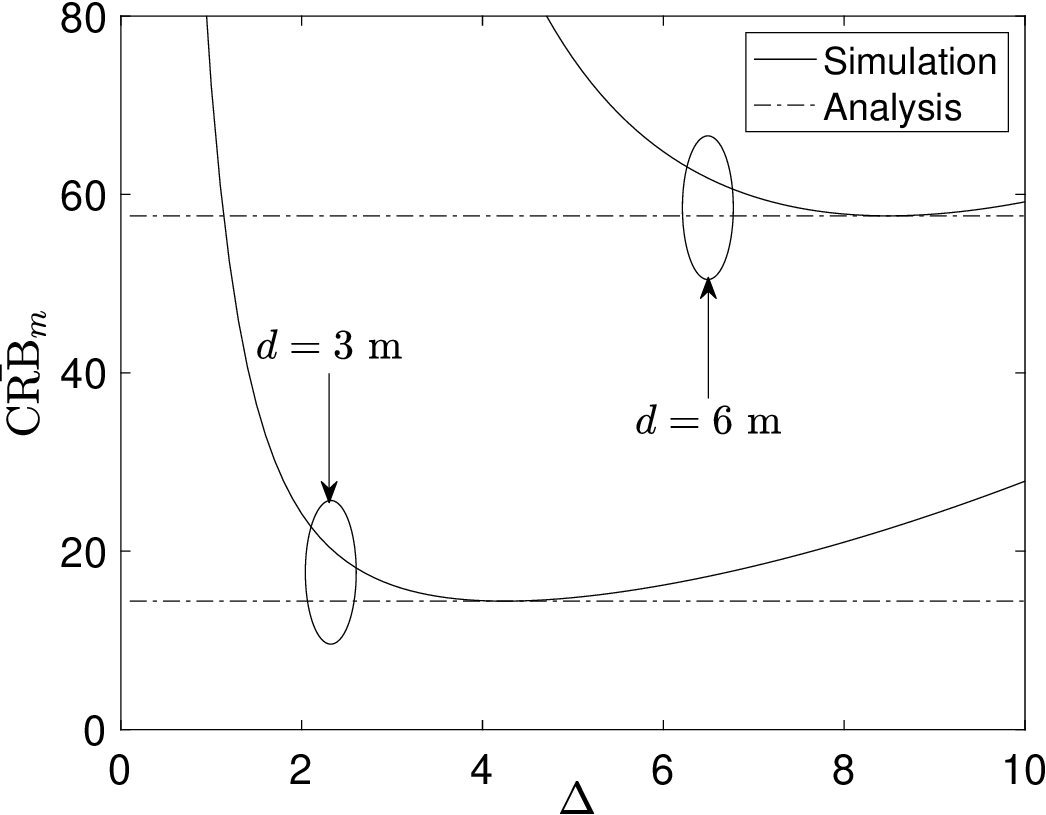, width=0.33\textwidth, clip=}\vspace{-0.5em}
\caption{Impact of the antenna spacing on the CRLB.  $N=4$ pinching antennas are activated in a square-shape area with the antenna spacing being $\Delta$ and ${\rm U}_m$ located at the center of the area, where $N_{\rm WG}=2$. The analytical results are based on \eqref{optimal}. 
  \vspace{-1em}    }\label{fig3}   \vspace{-0.8em} 
\end{figure}

Recall that one key feature of pinching antennas is their reconfiguration capabilities, where the number and the locations of the antennas can be changed in a flexible manner. Fig. \ref{fig2} demonstrates how this reconfiguration feature can be used to achieve flexible user-centric positioning.  In particular, Figs. \ref{fig2a} and \ref{fig2b} show that by activating the pinching antennas close to the intended user locations, different focal points can be realized, which means that users close to these focal points can enjoy high positioning accuracy.  For the case where the pinching antennas are clustered close to a user, Fig. \ref{fig3} is provided to show the impact of the antenna spacing on the CRLB, where the accuracy of the analytical results developed in \eqref{optimal} is also verified. 

 \vspace{-0.8em} 
\section{Conclusions}
This letter investigated how the key features of pinching antennas can be used to support ISAC from the CRLB perspective. In particular, the CRLB achieved by pinching antennas was first derived and then compared to that of conventional antennas. The presented analytical and simulation results demonstrated that the use of pinching antennas can significantly reduce CRLB and, hence, enhance the sensing capability.  In addition, this letter showed that the low-cost and reconfigurability features of pinching antennas can be utilized to realize flexible user-centric positioning.
  \vspace{-0.8em}
\bibliographystyle{IEEEtran}
\bibliography{IEEEfull,trasfer}

\begin{thebibliography}{10}
\providecommand{\url}[1]{#1}
\csname url@samestyle\endcsname
\providecommand{\newblock}{\relax}
\providecommand{\bibinfo}[2]{#2}
\providecommand{\BIBentrySTDinterwordspacing}{\spaceskip=0pt\relax}
\providecommand{\BIBentryALTinterwordstretchfactor}{4}
\providecommand{\BIBentryALTinterwordspacing}{\spaceskip=\fontdimen2\font plus
\BIBentryALTinterwordstretchfactor\fontdimen3\font minus
  \fontdimen4\font\relax}
\providecommand{\BIBforeignlanguage}[2]{{%
\expandafter\ifx\csname l@#1\endcsname\relax
\typeout{** WARNING: IEEEtran.bst: No hyphenation pattern has been}%
\typeout{** loaded for the language `#1'. Using the pattern for}%
\typeout{** the default language instead.}%
\else
\language=\csname l@#1\endcsname
\fi
#2}}
\providecommand{\BIBdecl}{\relax}
\BIBdecl

\bibitem{pinching_antenna2}
A.~Fukuda, H.~Yamamoto, H.~Okazaki, Y.~Suzuki, and K.~Kawai, ``Pinching antenna
  - using a dielectric waveguide as an antenna,'' \emph{NTT DOCOMO Technical
  J.}, vol.~23, no.~3, pp. 5--12, Jan. 2022.

\bibitem{mypa}
Z.~Ding, R.~Schober, and H.~V. Poor, ``Flexible-antenna systems: A
  pinching-antenna perspective,'' \emph{IEEE Trans. Commun.}, (to appear in
  2025) Available on-line at arXiv:2412.02376.

\bibitem{myblockage}
Z.~Ding and H.~V. Poor, ``Los blockage in pinching-antenna systems: Curse or
  blessing?'' \emph{IEEE Wireless Commun. Lett.}, (submitted) Available on-line
  at arXiv:2503.08554.

\bibitem{Kaidiactivation}
K.~Wang, Z.~Ding, and R.~Schober, ``Antenna activation for {NOMA} assisted
  pinching-antenna systems,'' \emph{IEEE Commun. Lett.}, (to appear in 2025)
  Available on-line at arXiv:2412.13969.

\bibitem{Chongjunpa1}
C.~Ouyang, Z.~Wang, Y.~Liu, and Z.~Ding, ``Array gain for pinching-antenna
  systems ({PASS}),'' \emph{IEEE Commun. Lett.}, (submitted) Available on-line
  at arXiv:2501.05657.

\bibitem{Chongjunpa1x}
Z.~Wang, C.~Ouyang, X.~Mu, Y.~Liu, and Z.~Ding, ``Modeling and beamforming
  optimization for pinching-antenna systems,'' \emph{IEEE Trans. Wireless
  Commun.}, (submitted) Available on-line at arXiv:2502.05917.

\bibitem{yanqingpa}
Y.~Xu, Z.~Ding, and G.~Karagiannidis, ``Rate maximization for downlink
  pinching-antenna systems,'' \emph{IEEE Commun. Lett.}, (to appear in 2025)
  Available on-line at arXiv:2502.12629.

\bibitem{xidongpass1}
X.~Mu, G.~Zhu, and Y.~Liu, ``Pinching-antenna system {(PASS)}-enabled multicast
  communications,'' \emph{IEEE Trans. Commun.}, (submitted) Available on-line
  at arXiv:2502.16624.

\bibitem{yuanwei1x}
J.~Xiao, J.~Wang, and Y.~Liu, ``Channel estimation for pinching-antenna systems
  {(PASS)},'' \emph{IEEE Trans. Commun.}, (submitted) Available on-line at
  arXiv:2503.13268.

\bibitem{yuanwei2x}
------, ``Beam training for pinching-antenna systems {(PASS)},'' \emph{IEEE
  Trans. Wireless Commun.}, (submitted) Available on-line at arXiv:2502.05921.

\bibitem{luyangpass1}
X.~Xie, Y.~Lu, and Z.~Ding, ``Graph neural network enabled pinching antennas,''
  \emph{IEEE Commun. Lett.}, (submitted) Available on-line at arXiv:2502.05447.

\bibitem{yuanwei3x}
J.~Guo, Y.~Liu, and A.~Nallanathan, ``{GPASS}: Deep learning for beamforming in
  pinching-antenna systems {(PASS)},'' \emph{IEEE Commun. Lett.}, (submitted)
  Available on-line at arXiv:2502.01438.

\bibitem{Tegosuppa}
S.~A. Tegos, P.~D. Diamantoulakis, Z.~Ding, and G.~K. Karagiannidis, ``Minimum
  data rate maximization for uplink pinching-antenna systems,'' \emph{IEEE
  Wireless Commun. Lett.}, (to appear in 2025) Available on-line at
  arXiv:2412.13892.

\bibitem{yuanwei5x}
M.~Sun, C.~Ouyang, S.~Wu, and Y.~Liu, ``Physical layer security for
  pinching-antenna systems {(PASS)},'' \emph{IEEE Commun. Lett.}, (submitted)
  Available on-line at arXiv:2503.09075.

\bibitem{yarupass}
Y.~Qin, Y.~Fu, and H.~Zhang, ``Joint antenna position and transmit power
  optimization for pinching antenna-assisted {ISAC} systems,'' \emph{IEEE
  Commun. Lett.}, (submitted) Available on-line at arXiv:2503.12872.

\bibitem{9737357}
F.~Liu, Y.~Cui, C.~Masouros, J.~Xu, T.~X. Han, Y.~C. Eldar, and S.~Buzzi,
  ``Integrated sensing and communications: Toward dual-functional wireless
  networks for {6G} and beyond,'' \emph{IEEE J. Sel. Areas Commun.}, vol.~40,
  no.~6, pp. 1728--1767, 2022.

\bibitem{4753258}
T.~Jia and R.~M. Buehrer, ``A new cramer-rao lower bound for {TOA}-based
  localization,'' in \emph{Proc. Military Commun. Conf.. (MILCOM 2008)}, Nov.
  2008, pp. 1--5.

\end{thebibliography}
  \end{document}